# Selectivity in tip-induced skeletal editing via heteroatom substitution

Shantanu Mishra,[1*] Rasmus Svensson,[1,2#] Valentina Malave,[3#] Florian Albrecht,[4] Manuel Vilas-Varela,[3] Henrik Grönbeck,[1,2] Leo Gross[4*] and Diego Peña[3,5*]

[1]Department of Physics and Astronomy, Chalmers University of Technology, 412 96 Göteborg, Sweden

[2]Competence Centre for Catalysis, Chalmers University of Technology, 412 96 Göteborg, Sweden

[3]Center for Research in Biological Chemistry and Molecular Materials, and Department of Organic Chemistry, University of Santiago de Compostela, 15782 Santiago de Compostela, Spain

[4]IBM Research Europe – Zurich, 8803 Rüschlikon, Switzerland

[5]Oportunius, Galician Innovation Agency, 15702 Santiago de Compostela, Spain

[#]R.S. and V.M. contributed equally.

[*]E-mail: shantanu.mishra@chalmers.se, LGR@zurich.ibm.com and diego.pena@usc.es.

**Abstract:** Skeletal editing enables precise structural modifications of molecules at late stages of a synthetic sequence, with applications in drug discovery and materials science. We recently demonstrated skeletal editing on the single-molecule scale. Voltage pulses applied by the tip of a scanning probe microscope to an oxygen-containing seven-membered heterocycle led to both oxygen deletion and ring-contraction rearrangement reactions. An open question is whether selective skeletal editing of a heterocyclic core can be achieved by an appropriate choice of the heteroatom. Here, we show that tip-induced reactions of an analogous sulfur-containing seven-membered ring results in sulfur deletion in virtually all cases. Our results demonstrate that the combination of tip-induced chemistry and heteroatom selection in the molecular design is a powerful strategy for single-molecule skeletal editing, with the potential to enable diverse structural transformations of heterocyclic frameworks.

**Scheme 1. (a) Illustration of skeletal editing.[a] (b, c) Schemes showing the major products formed from tip-induced skeletal editing of DNO (b) and DNT (c).**

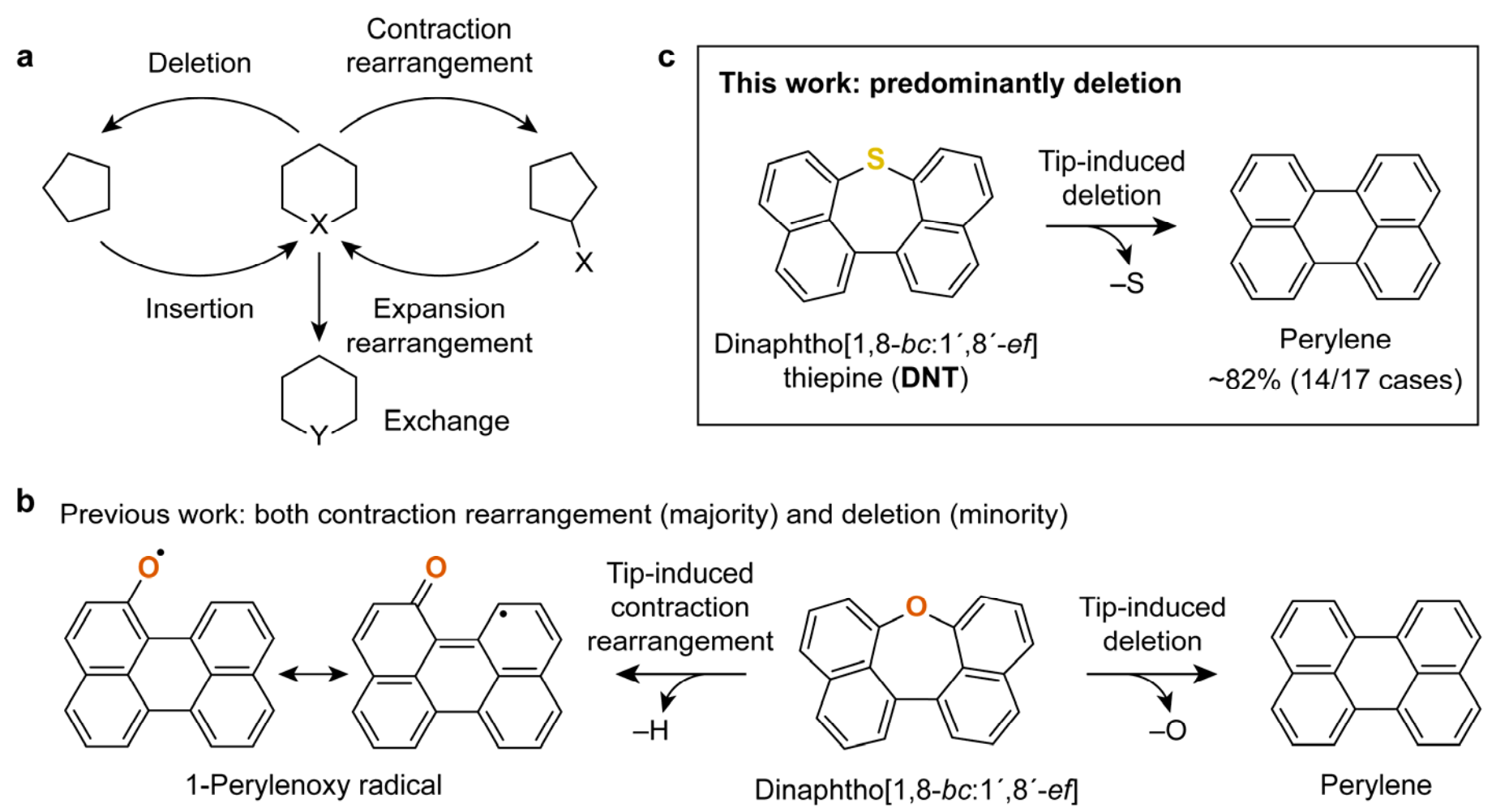

[a]X and Y denote individual atoms.

Skeletal editing refers to the late-stage modification of a molecular skeleton through insertion, deletion or exchange of individual atoms (Scheme 1a). The field has received considerable attention as a means of reducing the need for *de novo* synthesis, and precise structural transformation of molecules,[1,2] with potential applications in drug discovery, green chemistry, and materials science.[3–5] In particular, deletion of a strategically placed heteroatom from a carbon-based cyclic skeleton can provide access to diverse carbocyclic motifs. While conventionally performed via solution-phase chemistry (with demonstrated examples of boron,[6] nitrogen,[7–10] oxygen[11,12] and sulfur[13] deletion), extension of atom deletion to the single-molecule scale via tip-induced chemistry[14,15] could enable the synthesis of elusive molecules. In pursuit of single-molecule skeletal editing, we recently explored tip-induced reactions of individual dinaphtho[1,8-*bc*:1',8'-*ef*]oxepine (**DNO**) molecules adsorbed on NaCl (Scheme 1b).[16] Voltage pulses applied by the tip of a combined scanning tunneling microscope (STM) and atomic force microscope (AFM) to individual **DNO** molecules led to multiple reactions. Among all reacted molecules, nearly 50% corresponded to 1-perylenoxy radical resulting from ring-contraction rearrangement, whereas 33% corresponded to molecules generated via oxygen deletion, of which perylene was the major species (14% of all reacted molecules). The remaining molecules resulted from undesired side reactions. An open question is to what extent the selectivity of skeletal editing in a heterocyclic core can be directed toward a specific reaction by the choice of the heteroatom. In the present work we explore if heteroatom deletion can be favored over ring-contraction rearrangement by replacing oxygen with sulfur, hypothesizing that the lower C–S bond dissociation energy relative to the C–O bond would promote heteroatom deletion. It is shown that the tip-induced reactions of dinaphtho[1,8-*bc*:1',8'-*ef*]thiepine (**DNT**), which shares the same molecular structure as **DNO** but with sulfur replacing the oxygen atom, results in heteroatom deletion in virtually all cases, with perylene as the predominant product (Scheme 1c).

**DNT** was synthesized via solution-phase chemistry (Scheme S1, and Figures S1 and S2) and deposited onto a single-crystal Cu(111) surface partially covered by two monolayer-thick (denoted bilayer) NaCl films (Figure S3). Similar to **DNO**, individual molecules of **DNT** adsorbed on both Cu(111) and NaCl adopt a geometry where the sulfur atom points toward the surface. Figures 1a–c show STM and AFM images of a **DNT** molecule adsorbed on Cu(111). Related to the non-planar geometry of **DNT**, two diagonally-opposite benzenoid rings appear bright in AFM imaging due to their upward tilt. The remaining two benzenoid rings, which exhibit a downward tilt, appear notably dark, while the sulfur atom is not visible. Figures 1d–f show STM and AFM images of a **DNT** molecule adsorbed on NaCl, which exhibits stronger non-planarity because of weaker molecule-surface interaction on NaCl compared to Cu(111). The streaks in AFM images of **DNT**/NaCl (highlighted in Figure 1e) likely result from conformational switching of the molecule due to interaction with the tip. The adsorption geometry inferred from AFM imaging is in line with the density functional theory (DFT) calculation of the adsorption geometry of **DNT** (Figures 1g,h). Further measurements on **DNT** are shown in Figures S4 and S5.

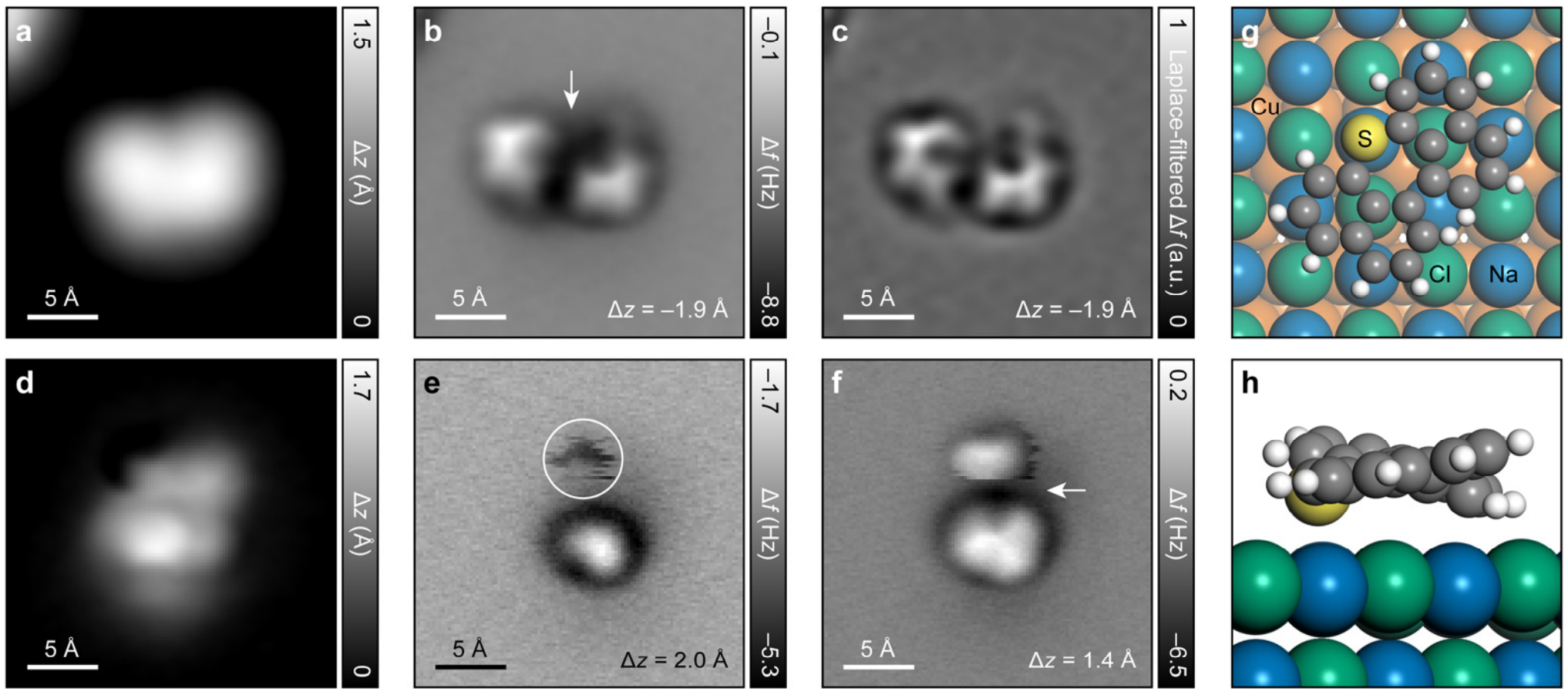

**Figure 1.** STM and AFM imaging of **DNT**. (a) STM image of a **DNT** molecule adsorbed on Cu(111). (b, c) AFM image (b) and the corresponding Laplace-filtered image (c) of the **DNT** molecule in (a). $\Delta z$ denotes the tip-height offset, with positive (negative) values indicating tip retraction (approach) from the set-point, $\Delta f$ denotes frequency shift, and a.u. denotes arbitrary units. (d) STM image of a **DNT** molecule adsorbed on bilayer NaCl. (e, f) AFM images of the **DNT** molecule in (d) taken at two different tip heights. The circle in (e) highlights streaks likely resulting from conformational switching of the molecule. The arrows in (b, f) indicate the locations of the sulfur atom. (g, h) Top and side views of the DFT-calculated adsorption geometry of **DNT** on bilayer NaCl/Cu(111). Carbon and hydrogen atoms are colored gray and white, respectively. Sodium, chlorine, copper and sulfur atoms are labeled in (g). Scanning parameters for STM images: $V = 0.2$ V, and $I = 0.30$ pA (a) and $I = 0.25$ pA (d). STM set-point for AFM images: $V = 0.2$ V, and $I = 0.50$ pA on bilayer NaCl (b, c) and $I = 0.25$ pA on Cu(111) (e, f). All images were acquired with the same tip.

We applied voltage pulses to individual **DNT** molecules adsorbed on NaCl to trigger skeletal editing reactions (we did not observe skeletal editing on Cu(111), see Figure S6). Briefly, we first positioned the tip above a **DNT** molecule at set-point conditions of $V = 0.2$ V and $I = 0.5$ pA ($V$ and $I$ denote bias voltage and tunneling current, respectively). Then, we opened the feedback loop and retracted the tip by 9.5–12.0 Å to limit the current (typically, $I <$ 30 pA at the final voltage). Finally, we increased the voltage to 4.7–5.0 V for 140 ms. If subsequent STM imaging indicated a change in the appearance of the molecule (Figure S7), we conducted AFM imaging to obtain the molecular structure. Upon application of voltage pulses to molecules adsorbed on a defect-free region on bilayer NaCl, we never found those molecules in a 200 Å × 200 Å scan frame. We presume that the molecules were picked up by the tip or displaced from NaCl islands onto Cu(111). Therefore, we applied voltage pulses to molecules adsorbed on bilayer NaCl at the step edges of third-layer NaCl islands, where the adsorption is stabilized. From 47 voltage pulses applied to 28 **DNT** molecules, we observed 17 molecules that underwent a reaction on the surface. In unsuccessful cases, the molecules were either displaced on NaCl without any reaction or not found. Sulfur deletion was the predominant reaction in successful cases (Figure 2), accounting for 16/17 reacted molecules (94%) and resulting in the generation of perylene (14/17 cases, ~82%) and, in a minority of cases, perylenyl radical (compound **1**, 2/17 cases). In one instance, we observed the generation of compound **2** resulting from a side reaction. Notably, there was no ring-contraction rearrangement observed for **DNT**, in contrast to **DNO** where it was the dominant reaction.

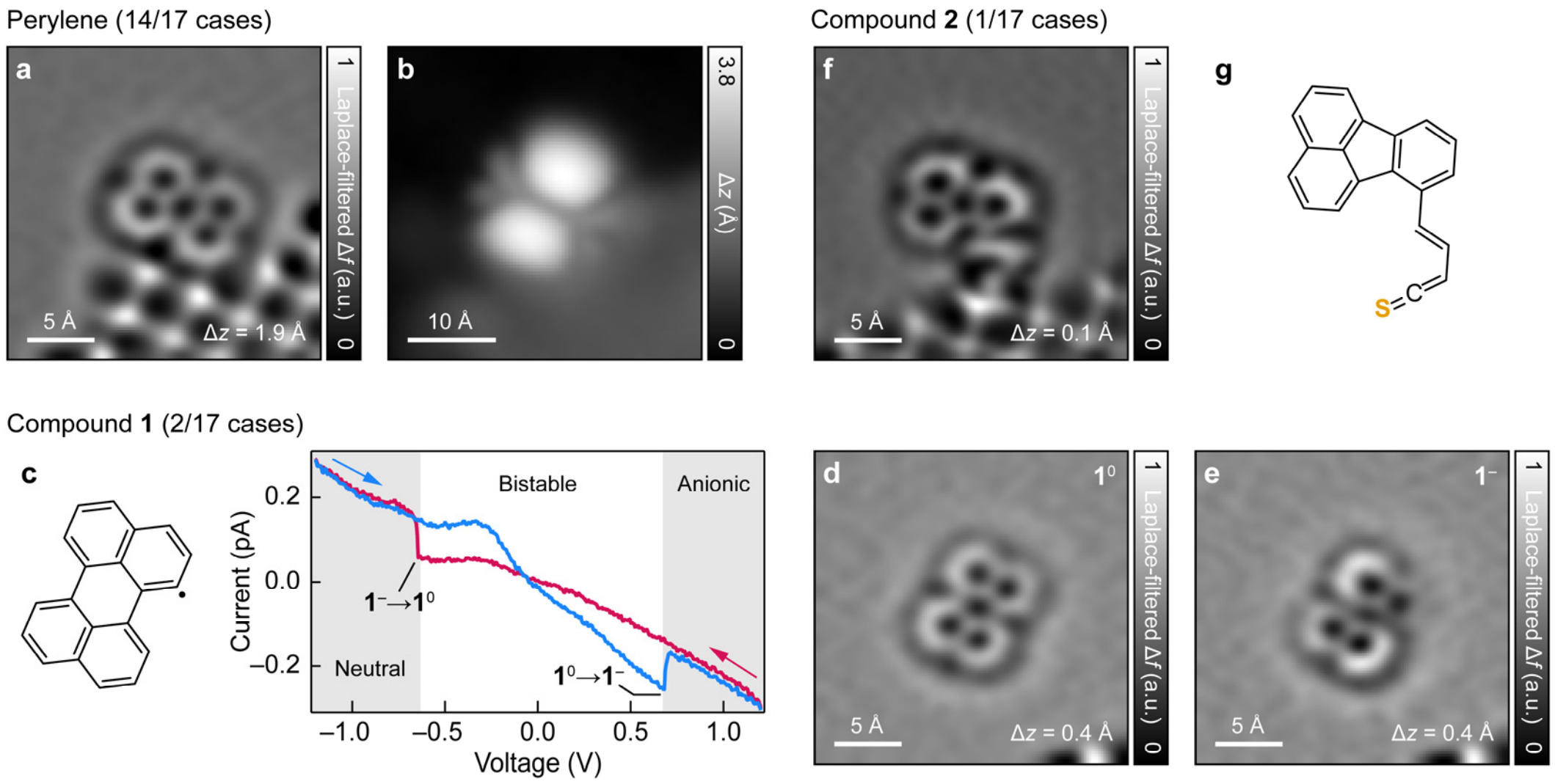

**Figure 2.** Structural and electronic characterization of reaction products on NaCl. (a) Laplace-filtered AFM image of a perylene molecule adsorbed at the step edge of a third-layer NaCl island. The bright and dark features in the third-layer NaCl island correspond to $Cl^-$ and $Na^+$ ions, respectively. (b) STM image of the perylene molecule in (a) showing its LUMO density ($V = 2.0$ V, $I = 0.3$ pA). (c) Constant-height $I(V)$ spectra acquired on **1** (right), revealing charge-state bistability. Open feedback parameters: $V = -1.2$ V, $I = 0.3$ pA; forward (backward) sweeps: –1.2 V to 1.2 V (1.2 V to –1.2 V). The molecule is in the neutral (anionic) state for $V < -0.65$ V ($V > 0.68$ V). The hysteretic behavior of the charging ($\mathbf{1^0}\rightarrow\mathbf{1^-}$) and discharging ($\mathbf{1^-}\rightarrow\mathbf{1^0}$) processes is related to the reorganization energy.[17,18] Within the hysteresis loop, **1** is bistable and can be imaged in both charge states. Also shown is the chemical structure of **1** (left). (d, e) Laplace filtered AFM images of $\mathbf{1^0}$ (left) and $\mathbf{1^-}$ (right). (f, g) Laplace-filtered AFM image of **2** adsorbed at the step edge of a third-layer NaCl island (left) and the proposed chemical structure of **2** (right). STM set-point for AFM images: $V = 0.2$ V and $I = 0.5$ pA on bilayer NaCl. The images in (d, e) were acquired with the same tip, while those in (a, b) and (f) were acquired with different tips.

Figure 2a shows an AFM image of a perylene molecule generated via tip-induced removal of a sulfur atom from **DNT** (see Figure S8 for AFM imaging of all perylene molecules generated in this study). STM imaging of the molecule at $V$ = 2 V revealed the lowest unoccupied molecular orbital (LUMO) density (Figure 2b), which agreed well with the calculated LUMO of perylene (Figure S9). This reinforced the assignment of the molecule as perylene. In a minority of cases, we also observed the generation of perylenyl radical (**1**), which exhibited charge-state bistability between neutral (denoted $\mathbf{1^0}$) and anionic (denoted $\mathbf{1^-}$) states (Figure 2c). AFM imaging revealed that whereas $\mathbf{1^0}$ is planar (Figure 2d), the trivalent carbon atom in $\mathbf{1^-}$ binds to the surface, leading to a non-planar adsorption geometry of $\mathbf{1^-}$ (Figure 2e). The assignment of charge states of **1** is supported by the absence (presence) of NaCl/Cu(111) interface-state scattering[19,20] by the neutral (charged) species (Figure S10). Finally, in one instance, we observed the generation of compound **2** resulting from a ring-opening and cyclization side reaction. Figures 2f,g show the AFM image and chemical structure of **2**, which consists of a fluoranthenyl core along with a functional group that we tentatively assign as buta-1,3-diene-1-thione.

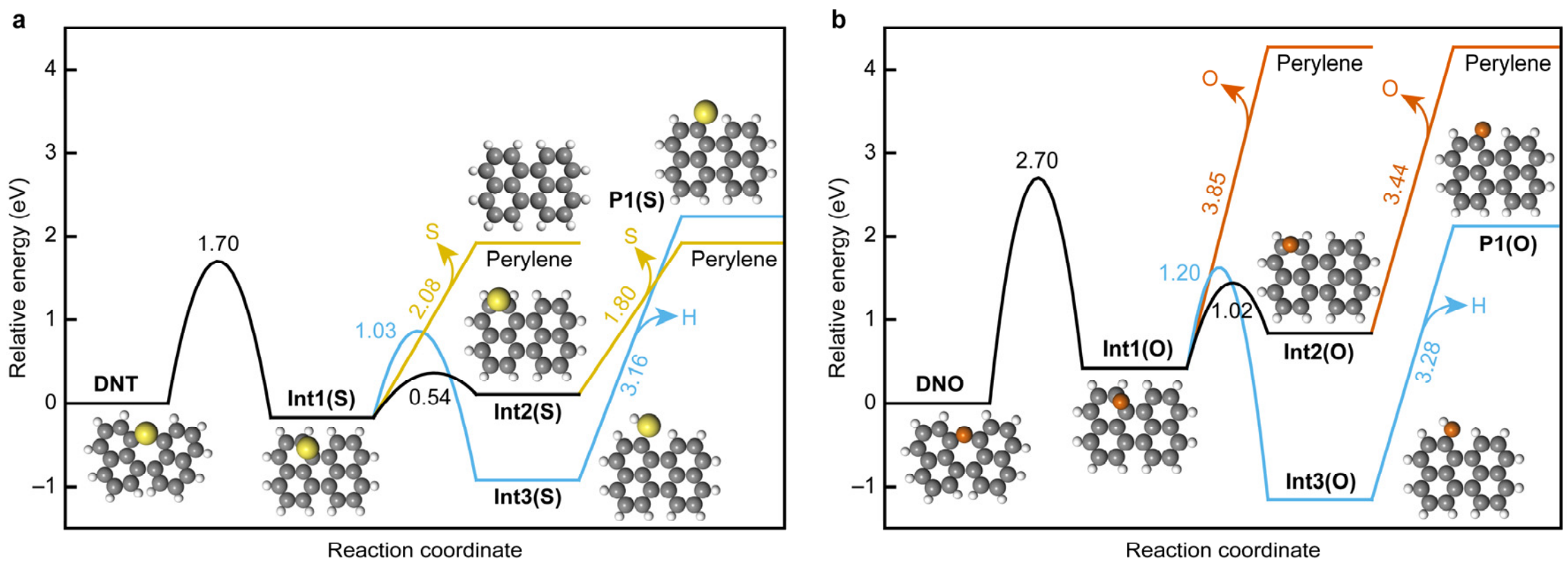

**Figure 3.** Gas-phase potential energy landscapes of skeletal editing of **DNT** (a) and **DNO** (b) obtained by DFT calculations. The numbers denote absolute gas-phase activation energies (in eV) of each reaction. For **Int1** and **Int2** → perylene and **Int3** → **P1**, energy differences coincide with activation energies. Also shown are the optimized gas-phase molecular geometries of precursors, intermediates, and products. Sulfur and oxygen atoms are colored yellow and orange, respectively. The calculations were performed assuming a neutral charge state of the reactant, intermediates, and products.

To elucidate the key experimental observations, namely, the predominance of atom deletion and the absence of ring-contraction rearrangement in skeletal editing of **DNT** relative to **DNO**, we performed gas-phase DFT calculations of the reaction landscape of **DNT** and compared the results with our previous findings for **DNO**.[16] Figures 3a,b show the reaction paths and energy barriers toward the generation of perylene and **P1**. Perylene results from atom deletion and is the predominant product of tip-induced reactions of **DNT**. **P1**, which results from ring-contraction rearrangement, was not observed in the present study but was the major product of tip-induced reactions of **DNO**. Starting from **DNT** (Figure 3a), the first step involves migration of the sulfur atom to a C–C bridge site and formation of a central six-membered ring. The formation of the resulting intermediate **Int1(S)** is associated with an activation energy $\Delta E^{\ddagger}$

= 1.70 eV, and the reaction is exothermic (the corresponding reaction **DNO** → **Int1(O)** is endothermic with $\Delta E^{\ddagger}$ = 2.70 eV; see Figure 3b). From **Int1(S)**, three reaction paths are possible, as previously observed for the case of **Int1(O)**. First, the sulfur atom can be directly eliminated, resulting in the generation of perylene. The associated $\Delta E^{\ddagger}$ for **Int1(S)** → perylene was found to be 2.08 eV, substantially lower than the $\Delta E^{\ddagger}$ = 3.85 eV for **Int1(O)** → perylene. Second, the sulfur atom can migrate to a neighboring bridge site, resulting in the intermediate **Int2(S)** ($\Delta E^{\ddagger}$ = 0.54 eV), from where it can again be eliminated to generate perylene, associated with $\Delta E^{\ddagger}$ = 1.80 eV (for **Int2(O)** → perylene, $\Delta E^{\ddagger}$ = 3.44 eV). Third, a rearrangement reaction **Int1** → **Int3** can occur, which exhibits similar $\Delta E^{\ddagger}$ values of 1.03 eV and 1.20 eV for the sulfur and oxygen cases, respectively, and is exothermic. From **Int3**, removal of the hydrogen atom bonded to the chalcogen results in the generation of the ring-contraction-rearrangement product **P1**, which exhibits similar $\Delta E^{\ddagger}$ values of 3.16 eV and 3.28 eV for the sulfur and oxygen cases, respectively. The substantially lowered $\Delta E^{\ddagger}$ (by more than 1.60 eV) for chalcogen elimination from **Int1** and **Int2** for sulfur compared to oxygen, along with the similar $\Delta E^{\ddagger}$ values for the reaction **Int3** → **P1** for sulfur and oxygen (the difference being only 0.12 eV), results in the high selectivity for atom deletion in **DNT**.

In conclusion, we demonstrate that selective skeletal editing of a heterocyclic core can be achieved through a judicious choice of the heteroatom. Combining single-molecule scanning probe experiments and density functional theory calculations, we show that tip-induced reactions of dinaphtho[1,8-*bc*:1',8'-*ef*]thiepine adsorbed on NaCl predominantly leads to atom deletion. This is in contrast to our previous study on the oxygen-containing analog dinaphtho[1,8-*bc*:1',8'-*ef*]oxepine, where both tip-induced ring-contraction rearrangement and atom deletion were observed as the majority and minority reactions, respectively. Our findings enable selective and atomically precise skeletal editing of heteroatom-containing carbon frameworks at the single-molecule scale.

**Acknowledgements**

This study has received funding from the European Research Council Synergy grant MolDAM (grant number 951519), the Spanish Agencia Estatal de Investigación (grant number PID2022-140845OB-C62), Xunta de Galicia (Centro de Investigación do Sistema Universitario de Galicia, 2023−2027, grant number ED431G 2023/03), the European Regional Development Fund, and the Swedish Research Council (grant number 2024-05250). The calculations were performed at PDC Center for High Performance Computing and the National Supercomputer Center (NSC) via a grant from the National Academic Infrastructure for Supercomputing in Sweden (NAISS).

**Associated Content**

**Supporting Information:** Experimental and theoretical methods, solution-phase synthesis of **DNT**, NMR characterization of **DNT**, STM and AFM data, and calculations.

**Notes**

The authors declare no competing financial interests.

# Supporting Information

## Selectivity in tip-induced skeletal editing via heteroatom substitution

Shantanu Mishra,[1] Rasmus Svensson,[1,2] Valentina Malave,[3] Florian Albrecht,[4] Manuel Vilas-Varela,[3] Henrik Grönbeck,[1,2] Leo Gross[4] and Diego Peña[3,5]

[1]Department of Physics and Astronomy, Chalmers University of Technology, 412 96 Göteborg, Sweden

[2]Competence Centre for Catalysis, Chalmers University of Technology, 412 96 Göteborg, Sweden

[3]Center for Research in Biological Chemistry and Molecular Materials, and Department of Organic Chemistry, University of Santiago de Compostela, 15782 Santiago de Compostela, Spain

[4]IBM Research Europe – Zurich, 8803 Rüschlikon, Switzerland

[5]Oportunius, Galician Innovation Agency, 15702 Santiago de Compostela, Spain

**Contents:**

# 1. Methods

## 1.1. Solution-phase synthesis and characterization

Starting materials were purchased reagent grade from TCI and Sigma-Aldrich and used without further purification. Binaphthalene **3** was synthesized following a reported procedure.[1] All reactions were performed in oven-dried glassware under an inert atmosphere of purified argon using Schlenk techniques. Thin-layer chromatography was performed on Silica Gel 60 F-254 plates (Merck). Column chromatography was performed on silica gel (40-60 µm). Nuclear magnetic resonance (NMR) spectra were recorded on a Varian Inova 300 spectrometer.

**Scheme S1. Synthesis of DNT.**

$Bu_6Sn_2S$
$Pd(PPh_3)_4$
Xylene
140 ºC, 16 h
Br
Br
S
**3**
**DNT**

Over a degassed mixture of **3** (500 mg, 1.21 mmol) and $Bu_6Sn_2S$ (630 µL, 1.21 mmol) in xylene (40 mL), the catalyst $Pd(PPh_3)_4$ (138 mg, 0.10 mmol) was added (Scheme S1). The resulting mixture was stirred at 140 ºC for 16 h. Then, the solvent was evaporated under reduced pressure, and the remaining residue was purified by column chromatography ($SiO_2$; hexane:$CH_2Cl_2$ 9:1), affording **DNT** (14 mg, 4%) as a yellow solid.

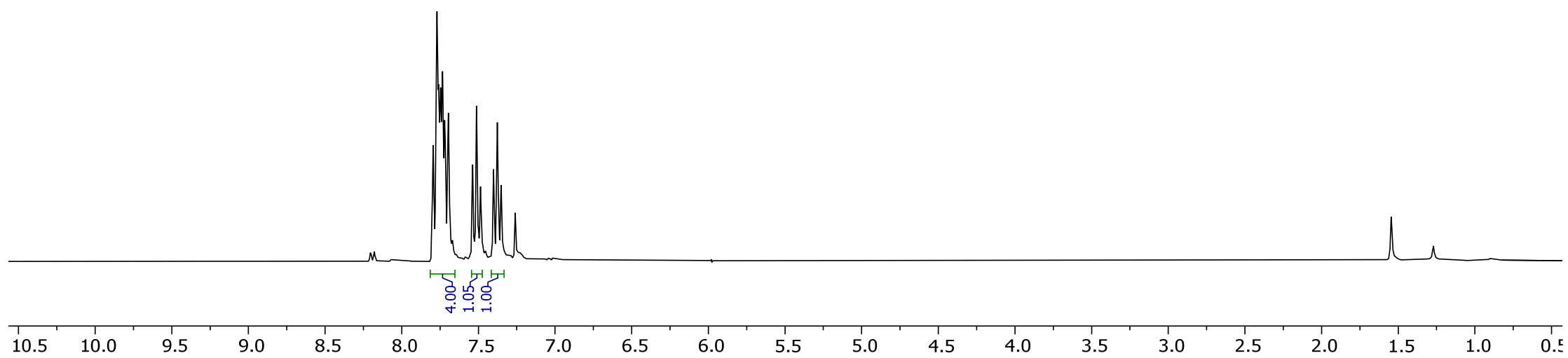

**Figure S1.** $^1$H NMR spectrum of **DNT**.

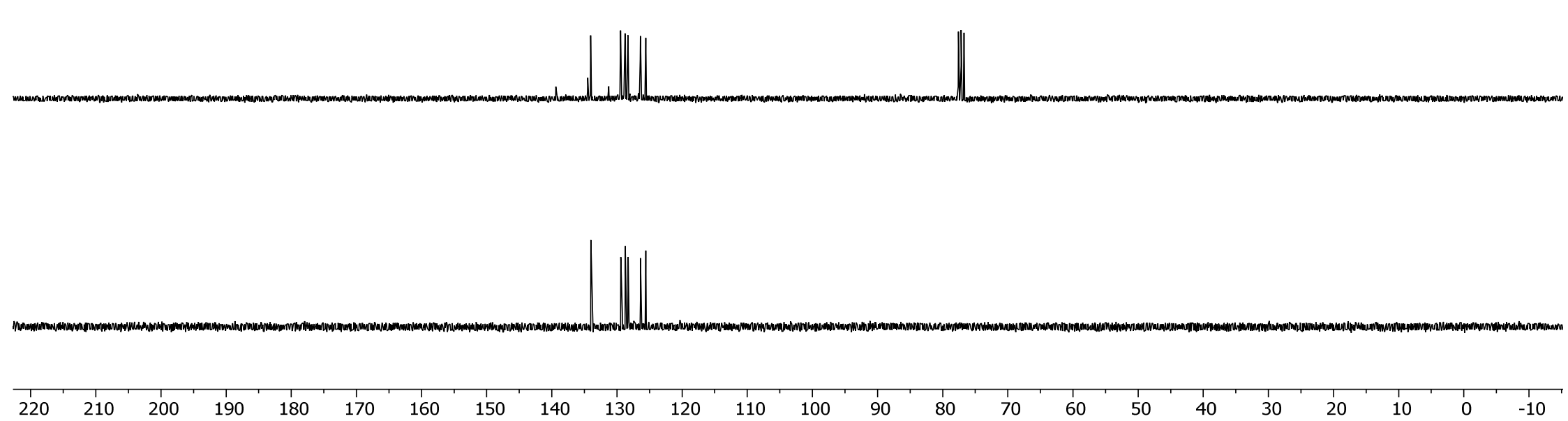

**Figure S2.** $^{13}$C NMR-DEPT spectra of **DNT** (DEPT: Distortionless Enhancement by Polarization Transfer).

### 1.2. Sample preparation and scanning probe microscopy measurements

STM and AFM measurements were performed in a home-built system operating at base pressures below $1\times10^{-10}$ mbar and a base temperature of 5 K. Bias voltages were applied to the sample with respect to the tip. Unless noted otherwise, all measurements were performed with copper-coated PtIr tips functionalized with a single carbon monoxide molecule at the tip apex. AFM imaging was performed in non-contact mode using a qPlus sensor.[2] The sensor was operated in frequency-modulation mode[3] with a constant oscillation amplitude of 0.5 Å. STM imaging was performed in constant-current mode, and AFM imaging was performed in constant-height mode at 0 V. STM and AFM images were post-processed using Gaussian low-pass filters.

The Cu(111) surface was prepared by multiple cycles of sputtering with $Ne^+$ ions and annealing up to 823 K. NaCl was thermally evaporated onto the Cu(111) surface held at 284 K. This protocol resulted in the growth of large and defect-free bilayer (100)-terminated NaCl films, with a minority of third-layer NaCl islands. The sample quality was checked by STM imaging prior to further preparation. Submonolayer coverage of **DNT** molecules on the surface was obtained by flashing an oxidized silicon wafer containing the molecules in front of the cold sample in the microscope. Carbon monoxide molecules were dosed from the gas phase onto the cold sample. The maximum sample temperature reached during deposition of **DNT** and carbon monoxide molecules was 16 K.

### 1.3. Density functional theory calculations

Density functional theory calculations were performed using the Vienna Ab Initio Simulation Package (VASP).[4–7] Interactions between core and valence electrons were described using the frozen core projector augmented wave (PAW) method.[8,9] The Kohn-Sham orbitals were expanded with plane waves, truncated at an energy cutoff of 450 eV. The considered valence electrons were $1s^1$ (H), $2s^2 2p^2$ (C), $2p^6 3s^1$ (Na), $3s^2 3p^4$ (S), $3s^2 3p^5$ (Cl) and $4s^1 3d^{10}$ (Cu). Exchange and correlation effects were described with the Perdew-Burke-Ernzerhof (PBE) functional.[10] For surface adsorptions, Grimme's D3 correction was included to account for the dispersion interactions.[11,12] The electronic structure was optimized until changes in the electronic energy and Kohn-Sham eigenvalues were below $1\times10^{-6}$ eV between succeeding iterations, whereas the nuclear structure was regarded as converged when the maximum nuclear force was below 0.03 eV/Å.

The lattice constant of NaCl was determined using a $p(1\times1)$ bilayer of NaCl (consisting of 4 Na and 4 Cl atoms), and optimized to be 5.526 Å. The lattice constant of bulk Cu was determined to be 3.570 Å, slightly lower than the experimental value of 3.610 Å.[13] The gas-phase species were optimized using a (32, 31, 30) Å vacuum box. Transition state calculations were performed using the climbing-image nudged elastic band method[14,15] using a (19, 18, 15) Å vacuum box. Molecular orbitals of the gas-phase molecules were calculated using Dmol$^3$ (ref.[16]). The Brillouin zones for all calculations were sampled by the Γ-point. The adsorption configuration of **DNT** was determined on a bilayer of $(5\sqrt{2} \times 4\sqrt{2})$R45° NaCl(100), supported on 2 layers of $(11 \times 5\sqrt{3})$rect. Cu(111). The optimized lattice constant for NaCl was used, whereas the Cu surface was strained to match the lattice of bilayer NaCl. The strains on Cu(111) were –0.50% and 1.11% in the two dimensions, respectively. In the calculations, the bottom layer of the strained Cu(111) was kept fixed, so as to emulate a bulk system.

## 2. Scanning probe data and calculations

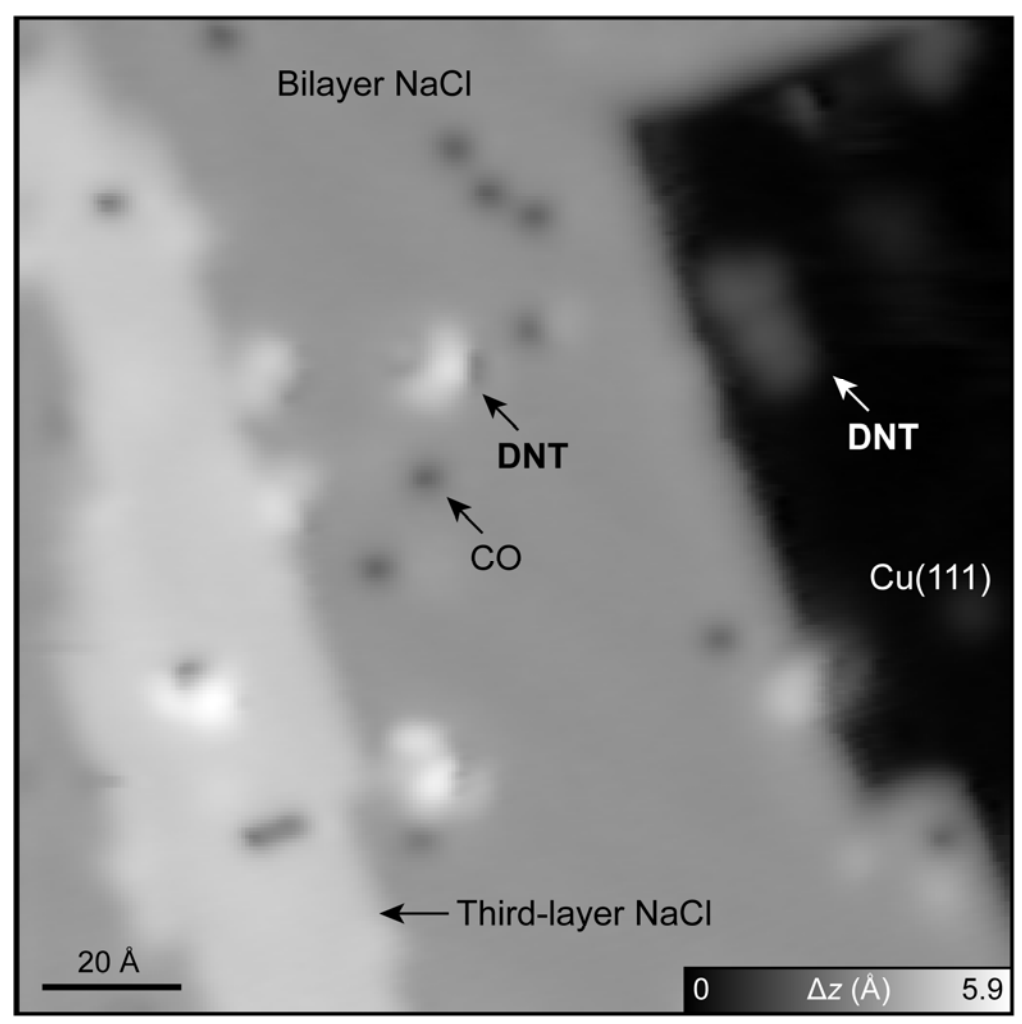

**Figure S3.** STM overview image of the sample. Scanning parameters: $V = 0.7$ V and $I = 0.3$ pA. Individual **DNT** molecules on NaCl and Cu(111), and co-adsorbed carbon monoxide (CO) molecules on NaCl are labeled. The image was acquired with a copper-coated metallic tip.

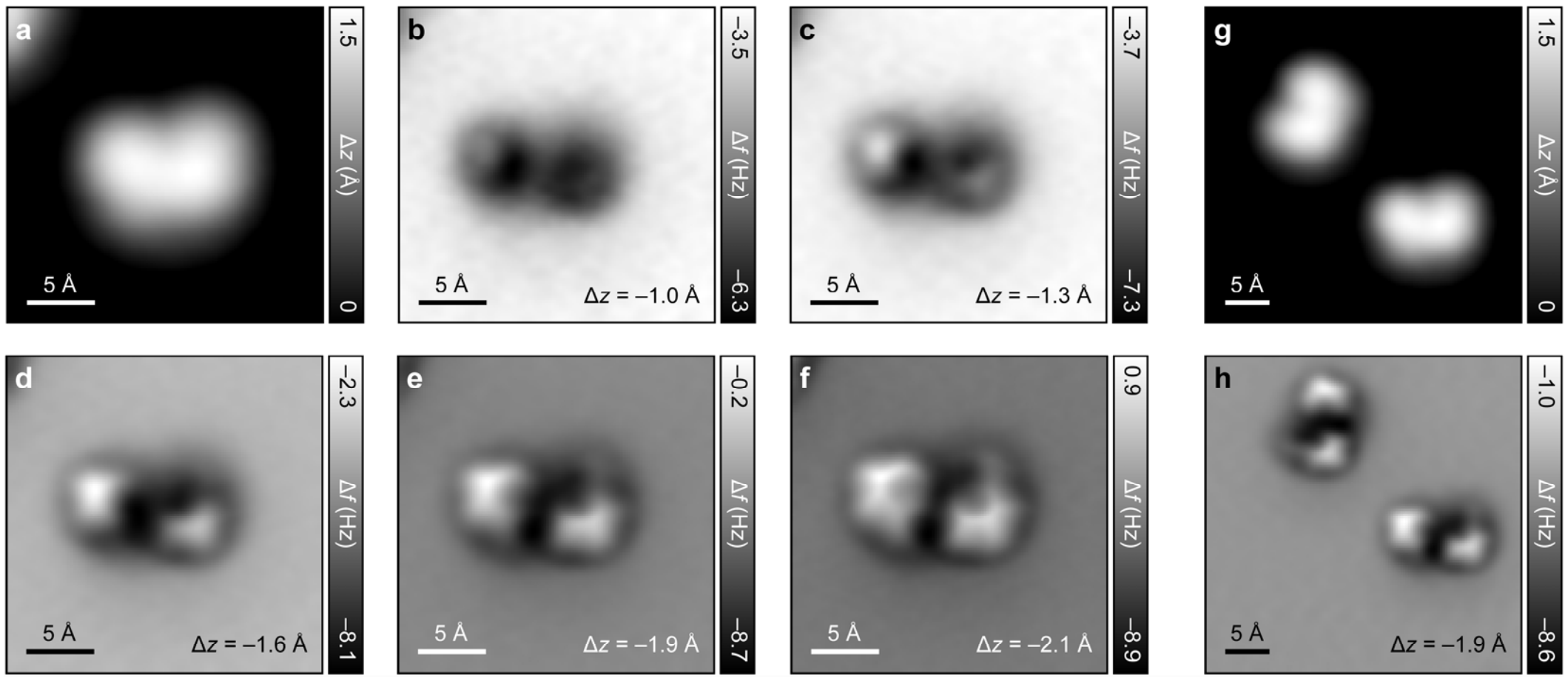

**Figure S4.** (a) STM image of a **DNT** molecule on Cu(111). (b–f) AFM images of the **DNT** molecule shown in (a) at different $\Delta z$ values. (g, h) STM (g) and AFM (h) images of two atropisomers of **DNT** on Cu(111). Scanning parameters for the STM image: $V = 0.2$ V and $I = 0.3$ pA. STM set-point for AFM images: $V = 0.2$ V and $I = 0.5$ pA on Cu(111). All images were acquired with the same tip.

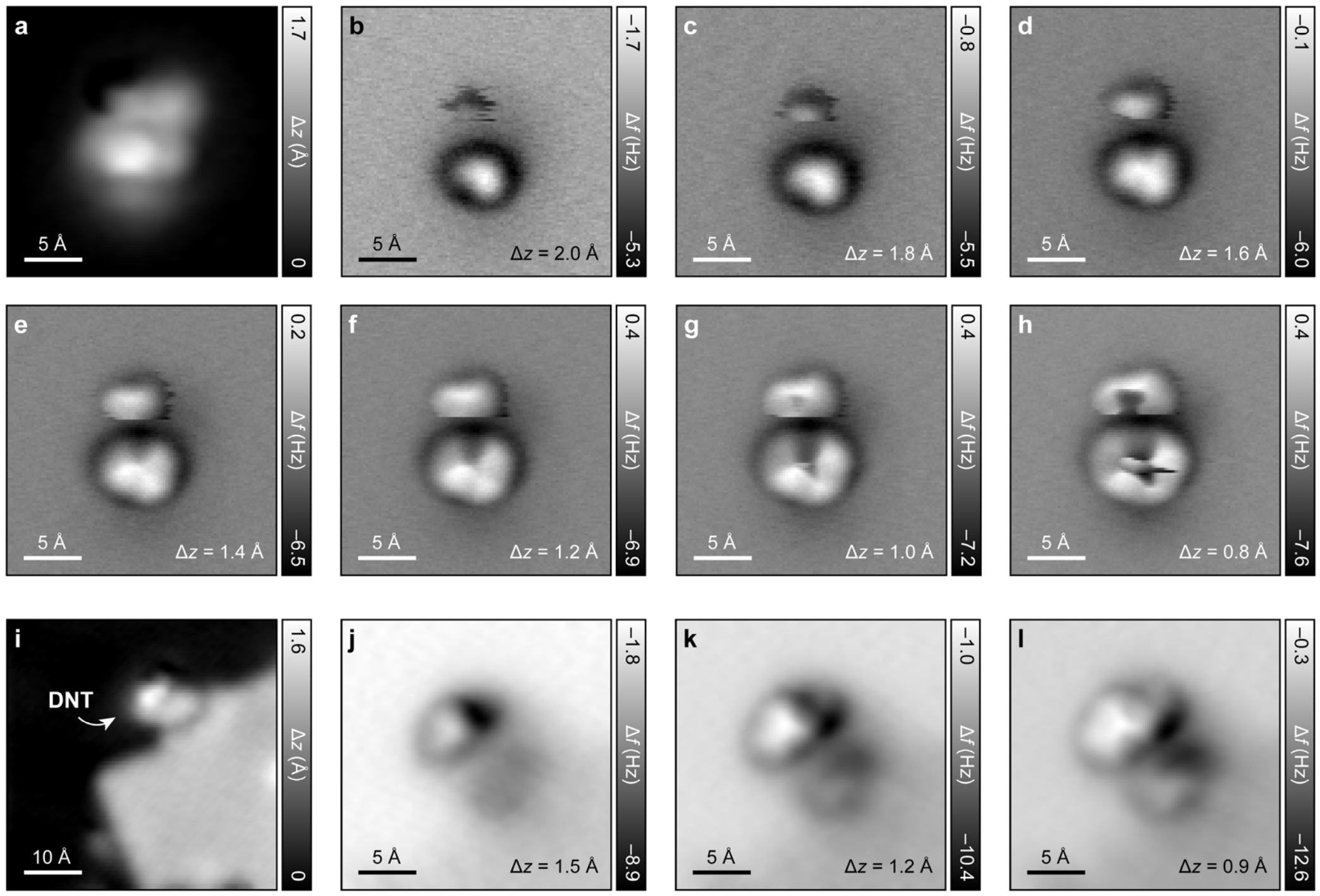

**Figure S5.** (a) STM image of a **DNT** molecule on NaCl. (b–h) AFM images of the **DNT** molecule shown in (a) at different $\Delta z$ values. (i) STM image of a **DNT** molecule adsorbed at the step edge of a third-layer NaCl island. (j–l) AFM images of the **DNT** molecule shown in (i) at different $\Delta z$ values. The adsorption of the molecule is stabilized at the step edge and conformational switching is suppressed, evidenced by the absence of streaks in AFM imaging. Scanning parameters for STM images: $V$ = 0.2 V, and $I$ = 0.25 pA (a) and $I$ = 0.50 pA (i). STM set-point for AFM images: $V$ = 0.2 V, and $I$ = 0.25 pA (b–h) and $I$ = 0.50 pA on NaCl. The images in (a–h) were acquired with the same tip, whereas those in (i–l) were acquired with a different tip.

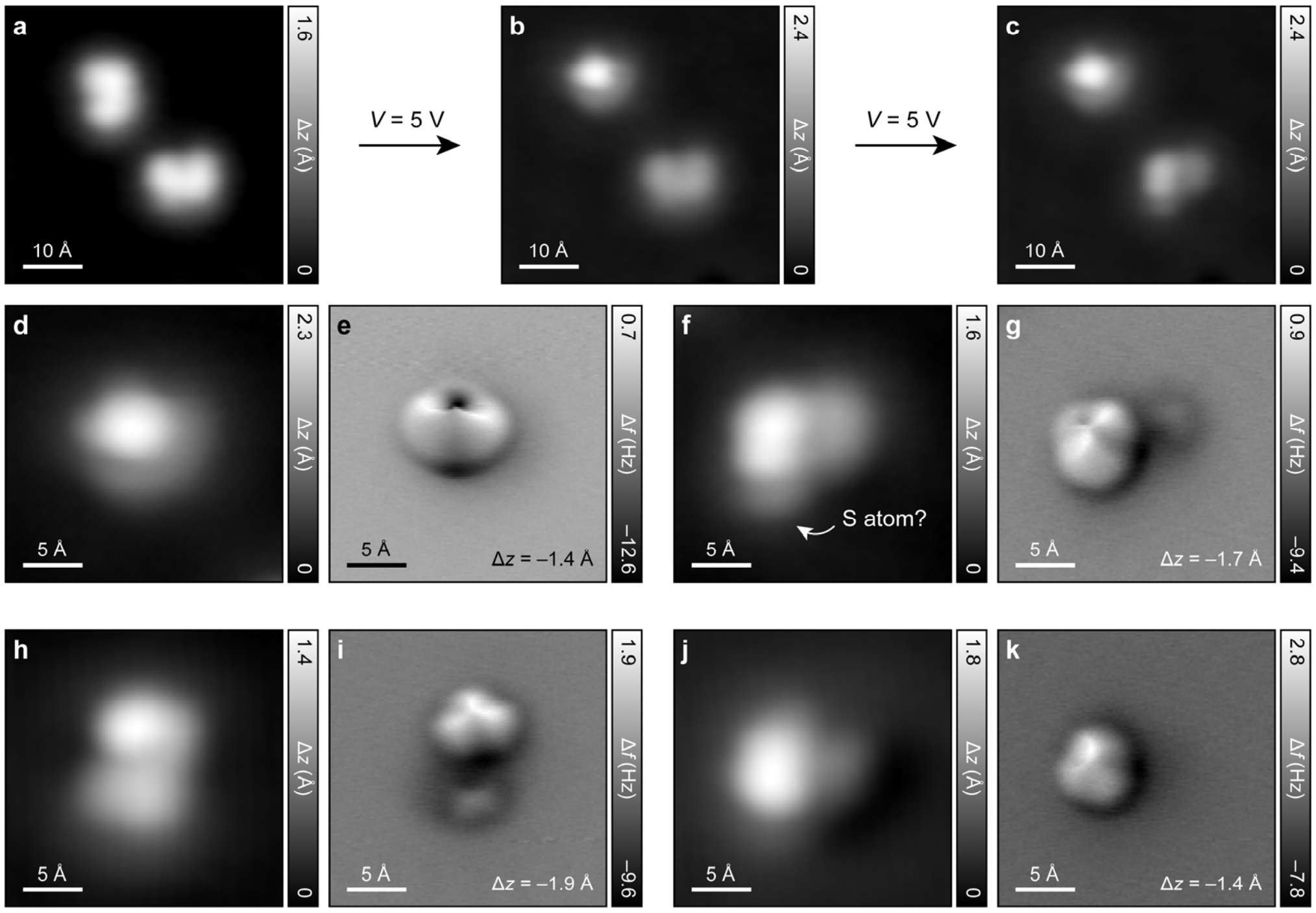

**Figure S6.** (a–c) STM images showing sequential tip-induced manipulation of **DNT** molecules on Cu(111): starting from two **DNT** molecules in (a), a voltage pulse of 5.0 V was first applied to the top molecule (b) and subsequently to the bottom molecule (c). (d–g) STM (d, f) and corresponding AFM (e, g) images of the molecules in (c). The species in (d, e) is difficult to identify. For the species in (f, g), two benzenoid rings (one of which exhibits a strong upward tilt) can be identified in (g). It is likely that the sulfur atom was abstracted from the molecule, resulting in the generation of a σ radical. The indicated feature in (f) could correspond to the detached sulfur atom. The strong non-planarity of the molecule probably results from the formation of C–Cu bond at the radical position(s). (h–k) STM (h, j) and AFM (i, k) images of two other species generated via application of voltage pulses of 5.0 V (h, i) and 5.1 V (j, k) to **DNT** molecules on Cu(111). The species in (h, i) and (j, k) exhibit different degrees of non-planarity. It is unclear from the images if the voltage pulses also triggered cleavage of C($sp^2$)–H bonds. Perylene molecules were not obtained on Cu(111). Scanning parameters for STM images: $V$ = 0.2 V and $I$ = 0.3 pA. STM set-point for AFM images: $V$ = 0.2 V and $I$ = 0.5 pA on NaCl. All images were acquired with the same tip.

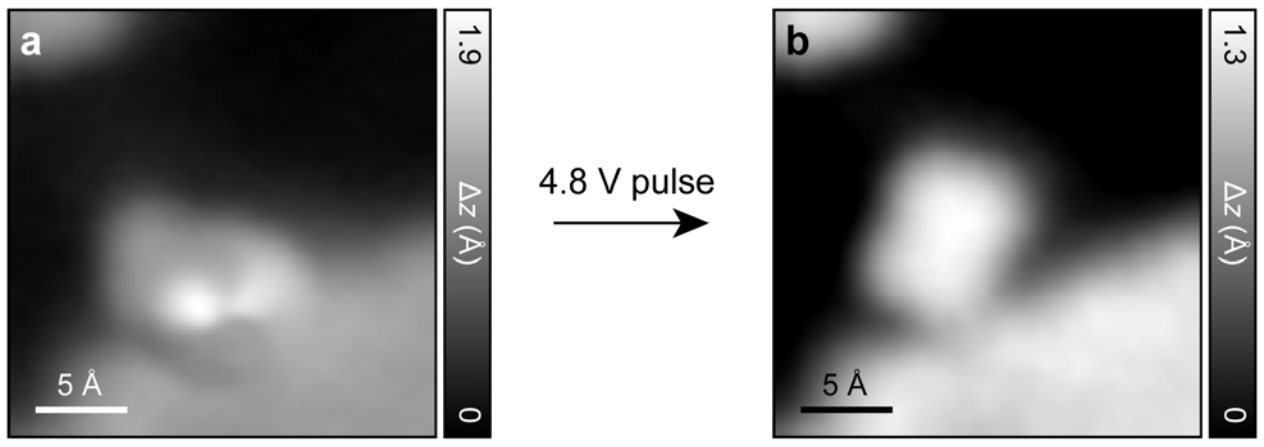

**Figure S7.** (a) STM image of a **DNT** molecule adsorbed at the step edge of a third-layer NaCl island. (b) STM image of the same area after application of a voltage pulse of 4.8 V to the molecule. The species in (b) corresponds to a perylene molecule, whose AFM image is shown in Figure S8a. Scanning parameters for STM images: $V = 0.2$ V and $I = 0.3$ pA. Before ramping the voltage to 4.8 V, the tip was retracted by $\Delta z = 10$ Å from the STM set-point of $V = 0.2$ V and $I = 0.5$ pA on top of the **DNT** molecule.

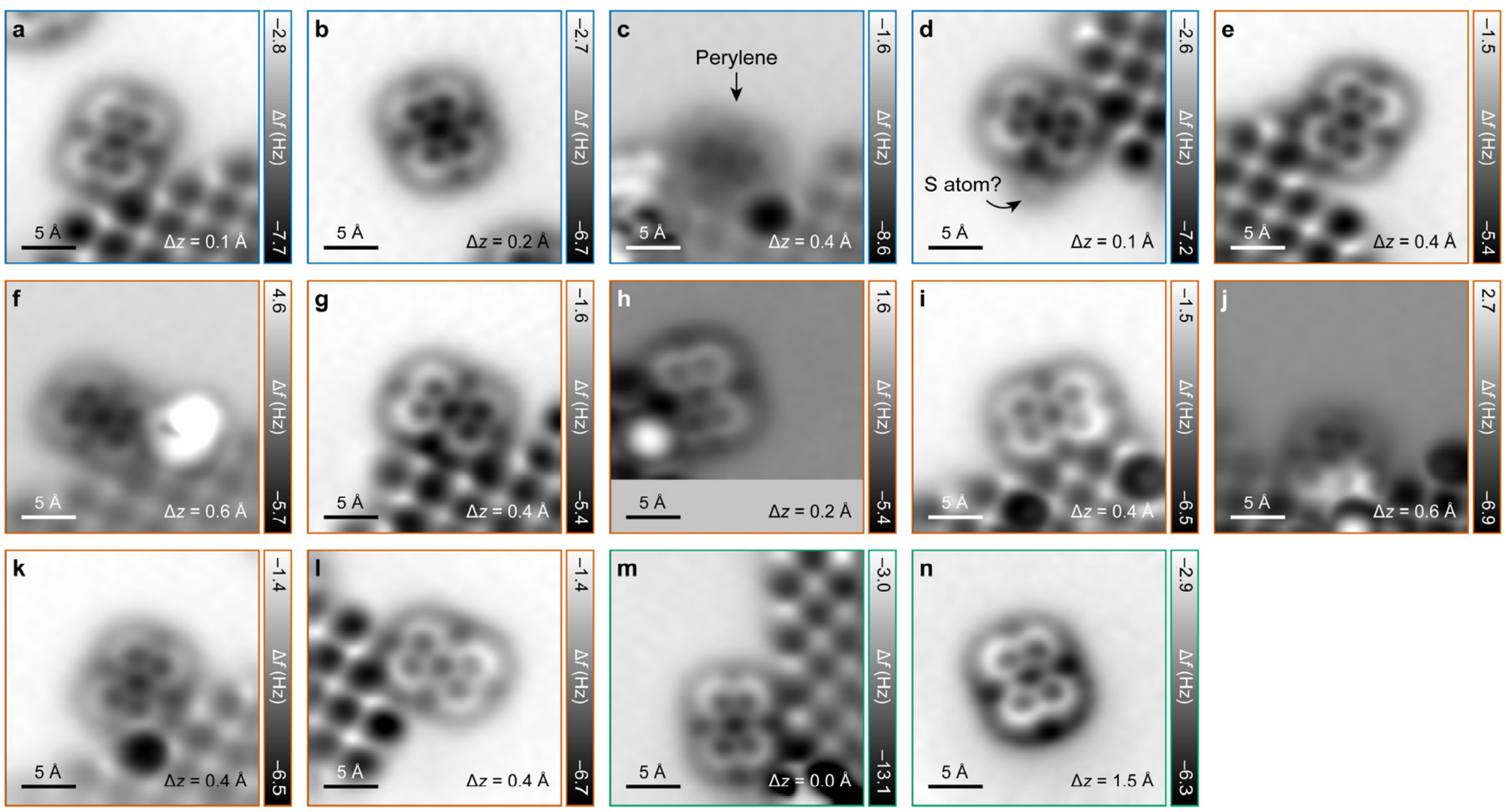

**Figure S8.** AFM images of all 14 perylene molecules generated via application of voltage pulses to **DNT** molecules on NaCl. The adsorbate indicated in (d) could correspond to a detached sulfur atom. This adsorbate was found to be negatively charged on the surface, inferred through the observation of NaCl/Cu(111) interface-state scattering by the adsorbate. The molecule in (j) is adsorbed at a copper step edge overgrown by NaCl. The molecule in (n) is adsorbed on a third-layer NaCl island. Images with the same frame colors were acquired with the same tip. STM set-point for AFM images: $V = 0.2$ V and $I = 0.5$ pA on NaCl.

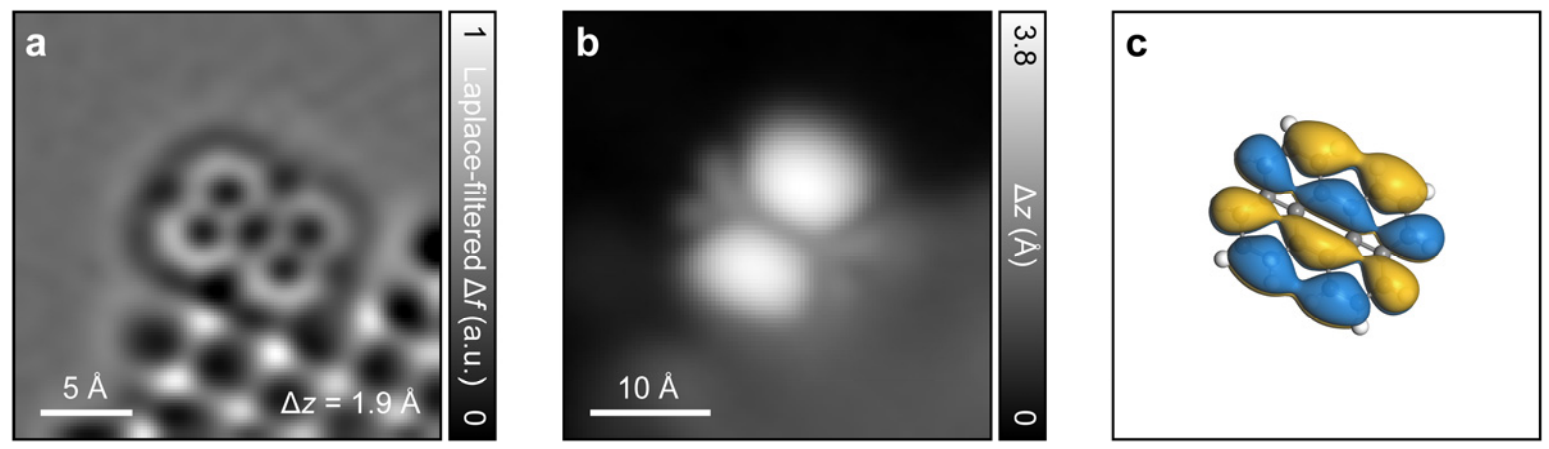

**Figure S9.** (a) Laplace-filtered AFM image of a perylene molecule generated on NaCl. (b) STM image of the perylene molecule in (a) showing its LUMO density ($V = 2$ V, $I = 0.3$ pA). Images (a) and (b) are also shown in Figure 2 of the main text. (c) The DFT-calculated LUMO of perylene (isosurface: $0.01 a_0^{-3/2}$, $a_0$ denotes the Bohr radius).

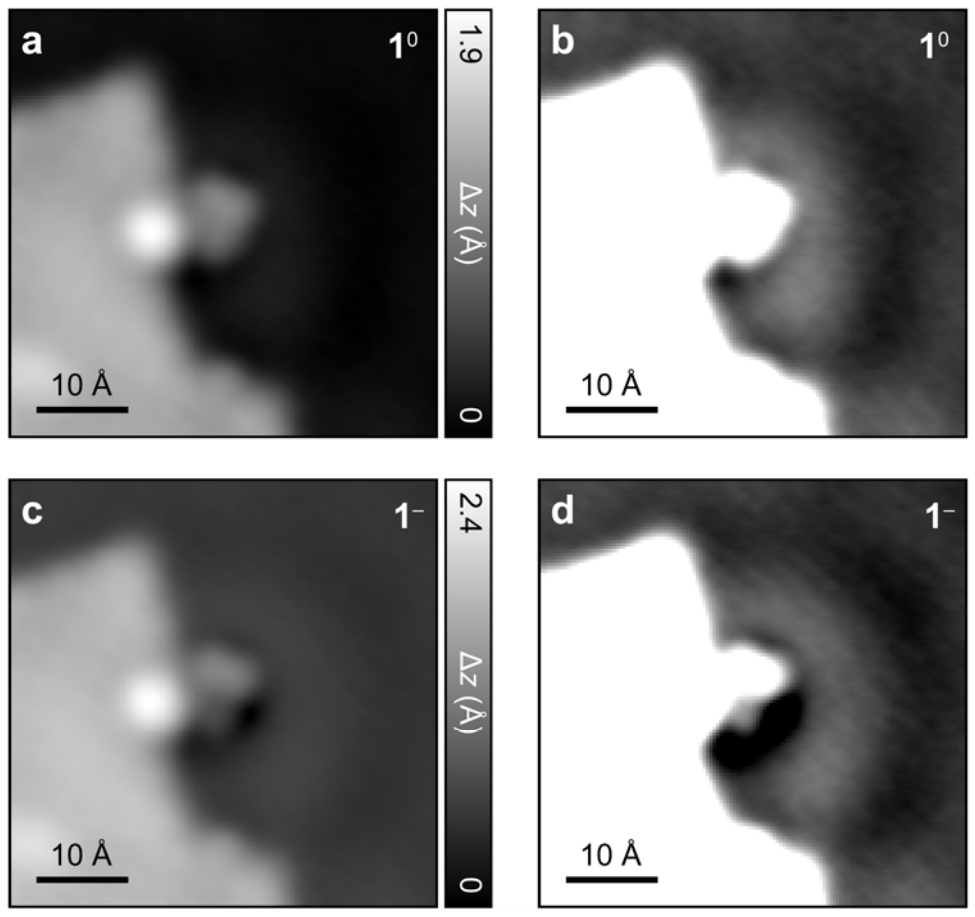

**Figure S10.** STM images of **1**, adsorbed at a third-layer NaCl step edge, in the neutral (a, b) and anionic (c, d) states. Image of **1** in each charge state is shown with two contrast levels. Strong NaCl/Cu(111) interface-state scattering[17,18] by $\mathbf{1}^-$ is observed (d), supporting its charged state. Scanning parameters: $V = 0.2$ V and $I = 0.3$ pA. The voltage in (a–d) is chosen to lie above the onset of NaCl/Cu(111) interface state at $V \sim -0.2$ V.[17]